\documentclass{emulateapj}

\newcommand{\xte}{{\it RXTE}}

\newcommand{\fpk}{$F_{\rm peak}$}
\newcommand{\fper}{$F_{\rm per}$}
\newcommand{\eps}{{\rm ergs\,s^{-1}}}

\newcommand{\epcs}{{\rm ergs\,cm^{-2}\,s^{-1}}}
\newcommand{\src}{4U~1636$-$536}
\newcommand{\osrc}{4U~1728$-$34}
\newcommand{\leddh}{$L_{\rm Edd,H}$}
\newcommand{\leddhe}{$L_{\rm Edd,He}$}

\newcommand{\fluen}{$E_{\rm b}$}
\newcommand{\tbb}{$T_{\rm bb}$}
\newcommand{\rbb}{$R_{\rm bb}$}

\newcommand{\burstnum}{123}			

\newcommand{\preburst}{40}
\newcommand{\nonpreburst}{78}
\newcommand{\highmean}{$6.4\times10^{-8}\ \epcs$}
\newcommand{\highsig}{7.6\%}
\newcommand{\highvar}{30}
\newcommand{\factor}{$1.69\pm0.13$}

\newcommand{\lastpublic}{2004 March 26}

\newcommand{\nonprerange}{ 0.21--$6.3\times10^{-8}\ \epcs$}
\newcommand{\prerange}{5.5--$7.4\times10^{-8}\ \epcs$}

\newcommand{\meandist}{$6.0\pm0.5$~kpc}
\newcommand{\mdnoerr}{6.0~kpc}
\newcommand{\maxdist}{6.6~kpc}
\newcommand{\conupper}{7.1~kpc}

\newcommand{\pcarspver}{10.1}
\newcommand{\lheasoftver}{5.3}
\newcommand{\lheasoftdate}{2003 November 17}

\slugcomment{Accepted by ApJ}

\shortauthors{Galloway et al.}
\shorttitle{Eddington-limited bursts from \src}

\begin{document}

\title{Eddington-limited X-ray Bursts as Distance Indicators. II. \\
Possible Compositional Effects in Bursts from \src\\}

\author{Duncan K. Galloway\altaffilmark{1,2},
  Dimitrios Psaltis\altaffilmark{3},
  Michael P.  Muno\altaffilmark{4,5}, and
  Deepto Chakrabarty\altaffilmark{6}}
\affil{Kavli Institute for Astrophysics and Space Research, 
  Massachusetts Institute of Technology, Cambridge, MA 02139}
\altaffiltext{1}{present address: School of Physics, University of Melbourne,
  Victoria, Australia}
\altaffiltext{2}{Centenary Fellow}
\altaffiltext{3}{present address: Department of Physics, University of Arizona,
  Tucson AZ}
\altaffiltext{4}{present address: Department of Physics and Astronomy,
  University of California, Los Angeles CA}
\altaffiltext{5}{Hubble Fellow}
\altaffiltext{6}{also Department of Physics, Massachusetts Institute of
Technology}
\addtocounter{footnote}{-6} 	
\email{D.Galloway@physics.unimelb.edu.au, dpsaltis@physics.arizona.edu,
  mmuno@astro.ucla.edu, deepto@space.mit.edu }

\begin{abstract}

We analyzed \burstnum\ thermonuclear (type-I) X-ray bursts
observed by the {\it Rossi X-ray Timing Explorer}
from the low-mass X-ray binary \src.
All but two of the \preburst\ radius-exansion bursts in this sample
reached peak fluxes which were 
normally distributed about a mean of \highmean, with a standard deviation of
\highsig. 
The remaining two radius-expansion bursts reached peak fluxes a factor of
\factor\ lower than this mean value;
as a consequence, the overall variation in the peak flux of the
radius-expansion bursts was a factor of $\approx2$.

This variation 
is comparable to the range of the Eddington limit 
between material with solar H-fraction ($X=0.7$) and pure He.
Such a variation may arise if, for the bright radius-expansion
bursts, most of the accreted H is eliminated either by steady hot CNO burning or
expelled in a radiatively-driven wind.
However, steady burning cannot
exhaust the accreted H for solar composition material within the typical
$\approx2$~hr burst recurrence time, nor can it result in sufficient
elemental stratification to allow selective ejection of the H only.
An additional stratification mechanism appears to be required to separate
the accreted elements and thus allow preferential ejection of the hydrogen.
We also observed non-radius expansion bursts that exceeded the
peak flux of the faintest radius expansion bursts. For these bursts the
accreted hydrogen must have been partly ejected or eliminated, but the
burst flux did not subsequently reach the (higher) Eddington limit for the
underlying He-rich material.

We found no evidence for a gap in the peak flux distribution between the
radius-expansion and non-radius expansion bursts, previously
observed in smaller samples.
Assuming that the faint radius-expansion bursts reached the Eddington
limit for H-rich material ($X\approx0.7$), and the brighter bursts the
limit for pure He ($X=0$), we estimate the distance to \src\ (for a canonical
neutron star with $M_{\rm NS}=1.4M_\odot$, $R_{\rm NS}=10$~km) to be
\meandist, or for $M_{\rm NS}=2M_\odot$ at most \conupper.
\end{abstract}


\section{INTRODUCTION}

Thermonuclear (type I) X-ray bursts are caused by unstable burning of
accreted matter on the surface of neutron stars in low-mass X-ray binary
(LMXB) systems \cite[see][for reviews]{lew93,bil98a}. Typical burst
profiles exibit short rise times between $\la 1$ to $10$~s, and decay
time scales from 10 to $\sim100$~s.  Model fits using a blackbody continuum
to X-ray spectra during the bursts provide evidence for an initial
rise in color temperature $T_{\rm bb}$, followed by a more gradual decrease
back to persistent levels. This is naturally interpreted as heating
resulting from thermonuclear ignition of surface fuel, followed by cooling of
the ashes once the available fuel is exhausted. The inferred blackbody
radius is around 10~km, consistent with expectations for a wide range of
neutron-star equations of state.
The time to achieve burst ignition depends primarily on the accretion
rate and the H-fraction, $X_0$, in the accreted material; the composition at
ignition, $X$ (as well as the temperature of the fuel layer) is modified by
steady hot-CNO H-burning between the bursts \cite[e.g.][]{fhm81}.
Such bursts have been
observed to date from more than 70 sources
\cite[for a recent catalog see][]{zand04b}.

If the energy from the burst is released sufficiently rapidly the 
flux may exceed the Eddington limit, which for a sufficiently distant
observer is \cite[]{lew93}
\begin{eqnarray}
  L_{\rm Edd,\infty} & = & \frac{8\pi G m_{\rm p} M_{\rm NS} c
  [1+(\alpha_{\rm T}T_{\rm e})^{0.86}]} {\sigma_{\rm T}(1+X)}
  \left(1-\frac{2GM_{\rm NS}}{Rc^2}\right)^{1/2} \nonumber \\ & = &
  3.5\times10^{38} \left(\frac{M_{\rm NS}}{1.4M_\odot}\right)
  \frac{1+(\alpha_{\rm T}T_{\rm e})^{0.86}}{1+X} \nonumber \\
& & \times\ \left(1-\frac{2GM_{\rm NS}}{Rc^2}\right)^{1/2}\ \eps
  \label{ledd}
\end{eqnarray}
where $M_{\rm NS}$ is the mass of the neutron star, $T_{\rm e}$ is the
effective temperature of the atmosphere, $\alpha_{\rm T}$ is a coefficient
parametrizing the $T$-dependence of the electron scattering opacity
\cite[$=2.2\times10^{-9}$~K$^{-1}$;][]{lew93}, 
$m_p$ is the mass of the proton, $\sigma_T$ the Thompson scattering
cross-section, and $X$ is the mass fraction of
hydrogen in the atmosphere ($\approx0.7$ for cosmic abundances).
The final factor
in parentheses represents the gravitational redshift due to the compact
nature of the neutron star, which also depends upon the height of the
emitting layer above the neutron star surface $R\ge R_{\rm NS}$.  Once the
Eddington limit is reached, the radiation 
forces due to the burst flux
are sufficient to lift the outer layers of the atmosphere above the
neutron star surface.
Thermonuclear bursts exhibiting photospheric radius-expansion (PRE) are
thus important because the peak flux can be estimated based on the
gravitational redshift and atmospheric composition. If the emission is
isotropic, such bursts represent a ``standard candle'', which can in
principle allow estimates of the distance to the source \cite[e.g.
see][]{kuul03a} or the
compactness of the neutron star \cite[e.g.][]{damen90}.

In this series of papers, we investigate empirically the assumption that
the peak flux of PRE bursts is constant for each burst source. In the
first paper, we found evidence for significant
variation in the peak flux of PRE bursts observed with
the {\it Rossi X-ray Timing Explorer} (\xte)  from \osrc\
\cite[][hereafter Paper A]{gal02c}.  The peak burst fluxes appeared to
vary steadily on a timescale of a few tens of days, which was similar to
the timescale at which the persistent X-ray flux was modulated. The
$\approx10$\% rms variation in peak burst flux was attributed to varying
degrees of reflection from a precessing accretion disk, with the burst
emission inferred to be intrinsically isotropic.
Here we present a study of the variation of the peak
fluxes of PRE bursts from \src, through analysis of the largest sample to
date, gathered from the 
available public data from observations by \xte.

\src\ ($l=332\fdg9$, $b=-4\fdg8$) is a well-studied LMXB, consisting of a
neutron star in a 3.8~h orbit with an 18th magnitude star, V801~Ara
(\citealt{1636orb}; see also \citealt{gil01}). The X-ray source exhibits a
variety of rapid time variability, including kHz quasi-periodic
oscillations \cite[]{wij97}, X-ray bursts, and burst oscillations at
579.3~Hz \cite[]{stroh98,stroh98b}.  Previous analyses of small numbers of
X-ray bursts observed from this source by various satellites revealed that
their peak fluxes appeared to be distributed bimodally in the most
part, with the PRE bursts reaching a peak flux a factor of 1.7 higher than
the brightest non-PRE burst \cite[]{inoue84b,lew87}.  
The measured peak fluxes of the PRE bursts were generally found to be
consistent from burst to burst \cite[e.g.,][]{eb87}.
The properties of the PRE bursts, along with
the claimed detection of redshifted absorption features in the burst
spectra \cite[]{waki84} have led to distance estimates between 6--7~kpc
\cite[e.g.][]{cs97}.

\section{RXTE OBSERVATIONS OF \src}
\label{obs}

The High-Energy Astrophysics Science Archive Research Centre (HEASARC;
{\url http://heasarc.gsfc.nasa.gov}) contains public \xte\/ observations
of \src\ dating from shortly after the launch of the satellite on 1995
December 30.  We extracted all the available public data, which at the
time of writing includes observations up to \lastpublic. We principally
used data from the Proportional Counter Array
\cite[PCA;][]{xte96} aboard \xte, which consists of five identical
gas-filled proportional counter units (PCUs) with a total effective area
of $\approx6000\ {\rm cm}^2$ and sensitivity to X-ray photons in the
2--60~keV range.  We have developed a pipeline processing system to
identify and download newly public data from around 70 known bursting
sources, including \src.
We generated 1-s binned lightcurves from each observation, and then
identified highly significant single-bin deviations from the mean count
rate as burst candidates. Each such candidate was visually inspected to
distinguish from other possible sources of abrupt count rate variation,
including PCUs being turned on or off, or PCU breakdowns. 

Once located, high time- and spectral resolution PCA data (where available)
covering each burst were processed to obtain
full-range spectra within intervals of 0.25--4~s (with the integration
time increasing as the burst count rate decays).
To take into account gradual variations in the PCA gain we
generated a response matrix for each burst using {\sc pcarsp} version
\pcarspver\footnote{We note that the geometric area of the PCUs was
changed for this release for improved consistency between PCUs and 
(e.g.) canonical models of calibration sources, particularly the Crab
pulsar and nebula.  These changes have the effect of reducing the measured
flux compared to analyses using previous versions of the response
generating tools by 12--14\%.  See
{\url http://lheawww.gsfc.nasa.gov/$\sim$keith/pca\_calibration\_draft.ps}
(Jahoda et al. 2005, ApJ, submitted) for more
details.}, which is part of {\sc lheasoft} release \lheasoftver\
(\lheasoftdate).
A persistent emission spectrum extracted from a (typically) 16~s interval
prior to the burst was used as the background; this approach is
well-established as a standard procedure in X-ray burst analysis (e.g.,
\citealt[]{kuul02a}, although see also \citealt[]{vpl86}).
We estimated the persistent flux, \fper, at the time of each burst by
fitting the background-subtracted spectrum averaged over the entire
observation (excluding the bursts) with a model consisting of an absorbed
blackbody and power law.

We fitted each time-resolved burst spectrum with a blackbody model
multiplied by a low-energy cutoff representing interstellar absorption
with fixed abundances. The initial fitting was performed with the
absorption column density $n_{\rm H}$ free to vary; subsequently, it was
fixed at the mean value measured over the entire burst for the final
results.  The 
flux at the peak of the burst did not vary significantly as a function of
$n_{\rm H}$.
The
bolometric flux at each timestep $t_i$ was calculated according to
\begin{eqnarray}
  F_{{\rm bol},i} & = & \sigma T_i^4 \left( \frac{R_{\rm NS}}{d} \right)_i^2 \nonumber \\
      & = & 1.0763\times10^{-11}\ T_{{\rm bb},i}^4 K_{{\rm bb},i}\ \epcs
\label{flux}
\end{eqnarray}
where $T_{\rm bb}$ is the color temperature,
$K_{\rm bb}=(R_{\rm bb,km}/d_{\rm 10 kpc})^2$ is the blackbody
normalisation, with $R_{\rm bb,km}$
the apparent
radius of the neutron star in km for a distance of $d_{\rm 10 kpc}\equiv d/(10\
{\rm kpc})$.  As a working definition, we considered that radius expansion
occurred when 1) the blackbody normalization $K_{\rm bb}$ reached a
(local) maximum close to the time of peak flux; 2) lower values of $K_{\rm
bb}$ were measured following the maximum, with the decrease significant to
a level of $4\sigma$ or more; and 3) there was evidence of a
significant (local) decrease in the fitted temperature $T_{\rm bb}$ at the
same time as the increase in $K_{\rm bb}$ (these criteria are identical to
those used in Paper A).
We measured the fluence \fluen\ by
integrating numerically over the measured values of $F_{{\rm bol},i}$,
extrapolating the derived exponential decay curve for the cases where the
burst emission lasted 
 \centerline{\epsfxsize=8.5cm\epsfbox{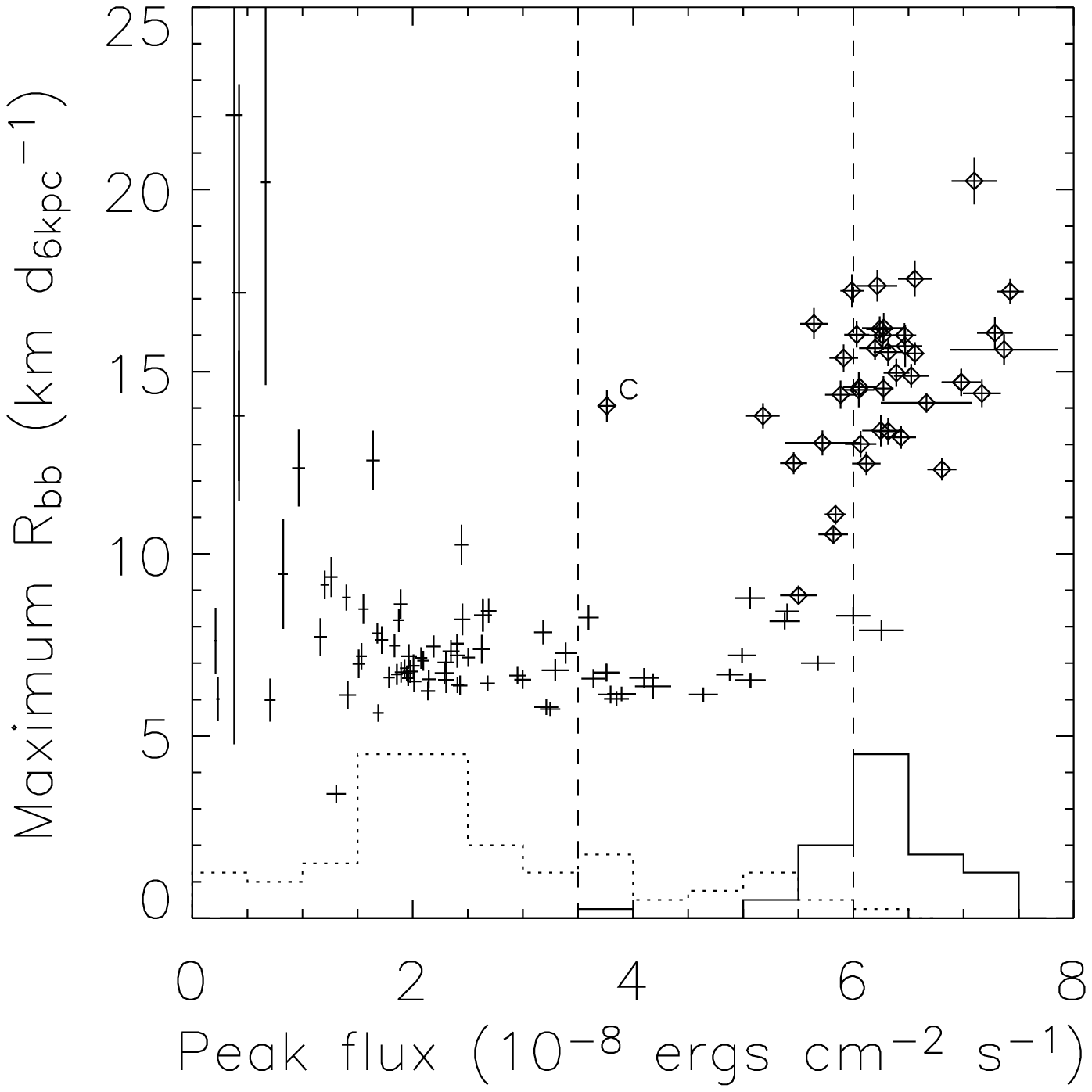}}
 \figcaption[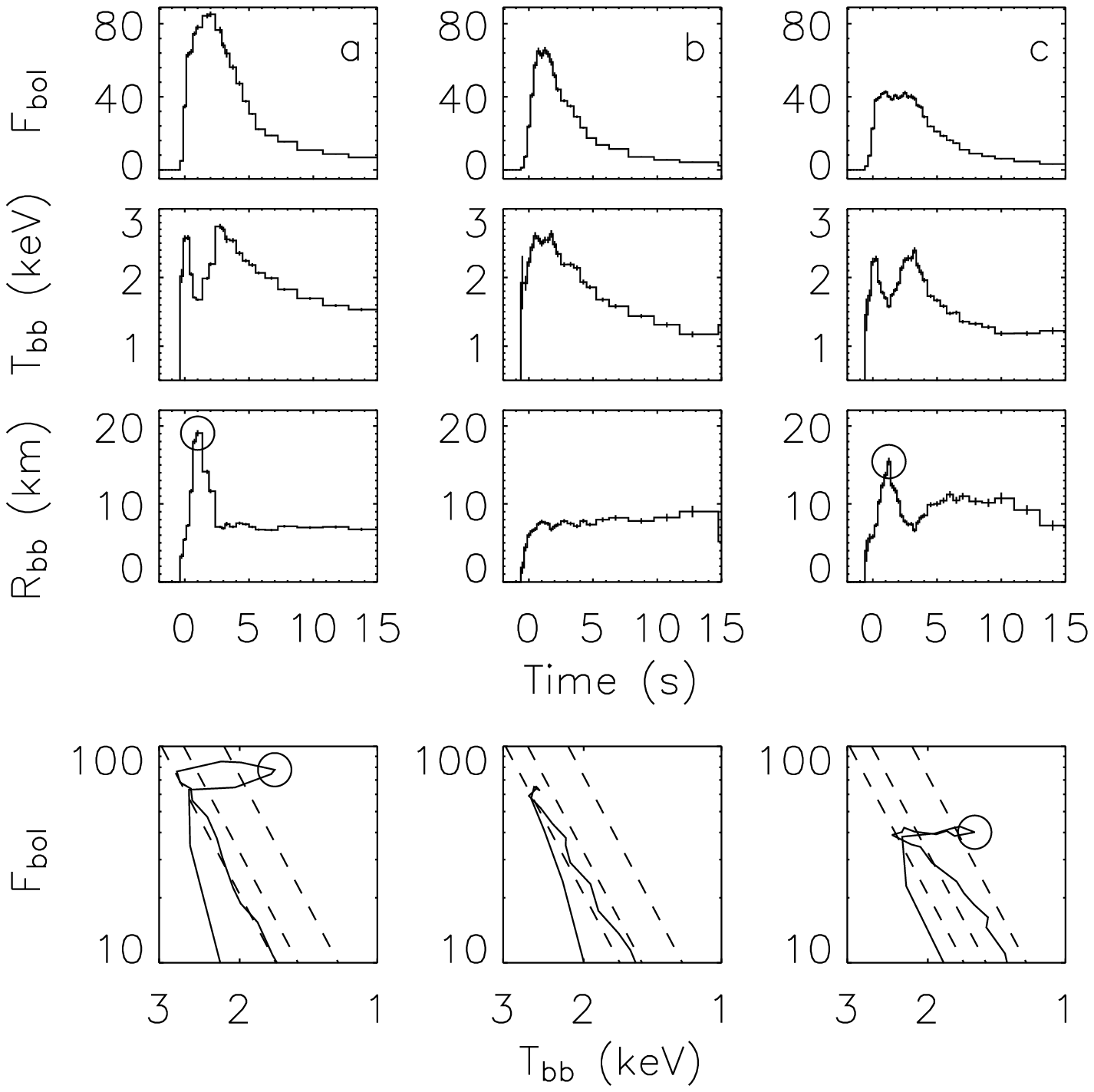]{Maximum blackbody radius (for a source distance of
\mdnoerr) in the rising phase of the burst as a function of the peak flux
for \nonpreburst\ non-radius expansion and \preburst\ radius expansion
bursts from \src. Error bars indicate the $1\sigma$ uncertainties on each
measurement; radius-expansion bursts are additionally indicated by diamond
symbols. Burst (c) is labeled (see \S\ref{loow}).
The dashed lines delineate the gap in the \fpk\ distribution inferred by
\cite{seh84}.
The histograms show the separate peak flux distributions of the 
non-radius expansion (dotted line) and radius expansion (solid
line) bursts. The largest bin for the two distributions contains 18
bursts.
 \label{pflux-hist} }
\noindent longer than the high-resolution spectral data
(typically 200~s).

\section{RESULTS}

The currently available public \xte\ data include \burstnum\ X-ray bursts
from \src.
The distribution of peak fluxes as a function of the presence or
absence of radius expansion is shown in Fig. \ref{pflux-hist}.  The
majority of the PRE bursts reached peak fluxes \fpk\ that were distributed
normally between \prerange, with a mean of \highmean\ and a standard
deviation of \highsig.  This variation was significantly greater than the
typical error on the flux ($\sim3$\%);
a $\chi^2$ calculation indicates that the hypothesis of a constant \fpk\ for
these bursts can be excluded down to a confidence level of $<10^{-16}$
(equivalent to $>8\sigma$).
The PRE bursts reached maximum \rbb\ approximately a factor of two larger
than the maximum achieved during the rise for the non-PRE bursts. The
weighted mean maximum radii were 
$14\pm2$ and $6.9\pm1.0\ {\rm km\,d_{6kpc}^{-1}}$,
respectively. The smallest maximum \rbb\ reached by a
burst which showed unambiguous indications of PRE was
$8.8\ {\rm km\,d_{6kpc}^{-1}}$.

We also observed \nonpreburst\ non-PRE bursts, with peak fluxes
distributed normally between \nonprerange.
We found several bursts (both PRE and non-PRE) which reached peak fluxes
between $\approx3.5$--$6\times10^{-8}\ \epcs$,
i.e. within the ``gap'' noted by \cite{seh84} and \cite{lew87}.
We also found two PRE bursts that reached a peak flux
of around
$3.7\times10^{-8}\ \epcs$,
which is lower than the maximum peak flux of the non-PRE bursts.
In the next section (\S\ref{loow}) we describe these bursts in
detail.

The \highsig\ spread in peak fluxes for the brighter (\fpk$\ga5\times10^{-8}\
\epcs$) PRE bursts is similar to that measured in \osrc\ (Paper A).
If the variation in the peak fluxes of the bright PRE bursts from
\src\ arises from the same mechanism, we might expect to find 
evidence for a connection with the persistent emission, as well as for
reprocessing of the burst flux (i.e. a correlation between the peak flux
and fluence).
While there are some indications for quasi-periodic variation at a period
of $\approx75$~d in the long-term All-Sky Monitor lightcurve, this
periodicity does not appear to be reflected by the peak burst fluxes.
We found no significant peaks in a Lomb-Scargle periodogram of the peak
fluxes as a function of time 
and no correlation between the peak PRE
burst flux and the fluence (as was also observed in \osrc).
We also found no evidence for modulation of the peak burst fluxes at the
3.8~hr orbital period (using the most recent published ephemeris;
\citealt{gil01}).

\subsection{Radius-expansion at low peak flux}
\label{loow}

We found two bursts that exhibited PRE but reached peak fluxes
significantly below the maximum peak flux of the non-PRE bursts. While one
of the bursts (on 1999 Sep 25 20:40:49 UT) exhibited a factor of $\sim2$
increase in radius and the archetypical flat-topped flux profile of a PRE
burst, the other (on 2000 Jan 22 04:43:48 UT) had much more modest
expansion and an overall less compelling case for PRE. The bursts reached
peak fluxes of
$3.76\times10^{-8}$ and $3.60\times10^{-8}\ \epcs$,
respectively. These peak fluxes were consistent to within the errors, and
were less than the mean for the remaining PRE bursts by a factor of
\factor\
(where the uncertainty arises primarily from the \highsig\
variation in the peak fluxes of the latter sample).
These bursts represent a
highly significant deviation from the normally-distributed peak fluxes of
the remaining PRE bursts, and lead to an overall variation of a factor of
$\approx2$ in the peak fluxes of PRE bursts from \src.

In Fig. \ref{example} we show the variation of spectral parameters
throughout a bright PRE burst, a non-PRE burst, and one of the fainter PRE 
bursts. 
The profiles
for the bright and faint PRE bursts (a, left panels and c, right panels)
were similar before and after the PRE
episode. However, for burst (c) the flux was approximately constant during
the PRE episode at around $4\times10^{-8}\ \epcs$, i.e. at around 60\% of that
reached at the peak of burst (a).
We note that the two faint PRE bursts were observed while the persistent
2.5--25~keV flux was unusually high, above $6\times10^{-9}\ \epcs$.  We
observed an insufficient number of bursts within this flux range to
determine conclusively whether the peak flux distribution differed from that
at lower persistent fluxes.

Both \rbb\ and \tbb\ measured at the time of maximum flux of the two faint
PRE bursts were, individually, within the ranges spanned by the brighter
PRE bursts.
Thus, the lower peak fluxes were not solely due to unusually low values of
either of these parameters.
We note that, as with previous observations \cite[]{seh84}, the radius
maximum during the majority of the PRE bursts was generally achieved prior
to the maximum measured flux.  Thus, the luminosity continued to increase
throughout the PRE episodes, as has been found on other sources
\cite[e.g.][]{gal03b}.
The two faint PRE bursts were notable exceptions; for those
bursts, the peak flux coincided with the peak radius.
Given the relatively wide field of view ($\approx1\arcdeg$) of \xte\ it is
conceivable that these fainter PRE bursts actually originated from a
previously unknown and more distant field source.  The detection of burst
oscillations at 580~Hz in the burst on 1999 September 25 (Fig.
\ref{example}c; \citealt{gil01,muno01}) effectively rules out this
possibility, at least for that burst.

A third burst, observed on 2002 January 15 14:08:16 UT, showed
evidence of a local radius maximum, but at an even lower peak flux than
burst (c).  During this burst the flux rose and decayed 
gradually, possibly with more than one local maxima.
The maximum radius reached was only somewhat below that of burst (c) and,
like that burst, was accompanied by a local minimum in \tbb. However, the
peak flux was below $2\times10^{-8}\ \epcs$, i.e. less than 25\% the peak flux
of burst (a); furthermore, the possible radius expansion episode occurred
well before the peak flux was reached, at a flux of only around
$1.3\times10^{-8}\ \epcs$. 
This burst was substantially different in character from the other two 
faint PRE bursts. It reached a comparable peak flux and was
observed at a similar persistent flux level as were four double-peaked
bursts, also observed by \xte.
These bursts (on 2001 September 5 08:15:04 UT, 2001 October 3 00:22:18 UT,
2002 January 8 12:22:44 UT and 2002 February 28 23:42:53 UT) exhibited
double peaks in the bolometric flux, as has been observed previously from
this source \cite[]{szt85}.
In contrast to the double-peaked bursts observed previously, in which the
first peak was consistently higher, 3 of the 4 double peaked bursts
observed by \xte\/ reached a higher flux during the second peak. The
relative fluxes of the two peaks varied from $\approx1$--3.
Two of the four double-peaked bursts exhibited a local maxima in the
radius (and a corresponding minimum in temperature) coincident or just
after the first flux peak, similar to what is generally interpreted as
evidence for PRE.
Because of this similarity, we conclude that the  2002 January 15 14:08:16
UT burst was not a genuine PRE burst, but instead arose from the same
phenomenon which gives rise to the double-peaked bursts.

\section{DISCUSSION}
\label{disc}

We have studied the peak fluxes of radius expansion bursts from \src\
and found the largest range yet seen in any LMXB.  While
the fractional variation of peak fluxes exceeded the expected range of the
Eddington limit between atmospheres with H at cosmic abundances
(\leddh, with $X\simeq0.7$) and with pure He \cite[\leddhe, $X=0$;
e.g.][]{kuul03a}, the distribution of peak fluxes was nearly bimodal,
with the mean peak flux for the majority of the PRE bursts being a
factor of \factor\ higher than the peak fluxes for the faint PRE
bursts.

In a previous investigation of X-ray bursts from \src\ with {\it
Tenma\/}, \cite{seh84} observed a gap in the distribution of peak
fluxes of bursts and identified the lower and upper boundaries of this
gap as the Eddington limits for H-rich \leddh\ and pure He fuel
\leddhe, respectively. According to their interpetation, the increase
in radius which was observed during the PRE bursts from \src\ was
accompanied by 
 \centerline{\epsfxsize=8.5cm\epsfbox{f2.eps}}
 \figcaption[example.eps]{Thermonuclear bursts from \src\ showing 
radius expansion over a wide range of peak fluxes. From left to right, we 
show a PRE burst with strong radius expansion, observed on 
1998 Aug 20 03:40:09 UT; 
a non-PRE burst observed on 2000 Jan 22 01:46:23 UT; 
and a faint PRE burst observed on 1999 September 25 20:40:49 UT. 
From top to bottom, the
panels show for each burst the time evolution of the bolometric flux
$F_{\rm bol}$ (in units of $10^{-9}\ \epcs$), the fitted blackbody
temperature $T_{\rm bb}$, the blackbody radius $R_{\rm bb}$ (assuming a
source distance of \mdnoerr),
and finally the path traced by the burst in $T_{\rm bb}$--$F_{\rm bol}$
space. The circles in the lower two panels in each column indicate the
time of maximal radius expansion, where present. The 3 dashed lines in
each of the bottom panels represent the expected curves for blackbody
emission from a neutron star with apparent radius 8, 10, and 15~km (moving
left to right, respectively).  Error bars represent the $1\sigma$
uncertainties.
 \label{example} }
\noindent ejection of an outer, H-rich layer, exposing the
underlying pure-He layer below. Bursts that never reached \leddh\ did not
exhibit PRE, while the characteristic flux limit for bursts that did
exceed \leddh\ switched to \leddhe, which is a factor of 1.7 higher.

In this paper, we reported the observation of two PRE bursts that were
apparently limited by \leddh\ instead of \leddhe. This appears to
confirm the hypothesis of \cite{seh84}. For those bursts, the
effective Eddington limit could have remained at \leddh\ if the burst
energy was insufficient to drive off the outer H-rich layer.  However,
we also found a significant overlap between the peak flux
distributions for the PRE and non-PRE bursts, without the gap found in
earlier samples.  Bursts with peak fluxes within the range
$\approx3.5$--$6\times10^{-8}\ \epcs$ were, however, relatively
infrequent.  This strongly suggests that the previously observed gap
was merely a consequence of small burst samples, specifically 12 for
the study of \cite{seh84}, and 27 for that of \cite{lew87}, with only
3 radius expansion bursts in each sample.

Four bursts observed by \xte\/ that peaked between the putative
\leddh\ and \leddhe\ values exhibited no evidence of PRE (an example
is shown in Fig. \ref{example}b).  In order for these bursts to have exceeded
\leddh, the majority of the accreted H must have been ejected, but
subsequently the flux must have remained below the effective Eddington
limit for the residual material. The fact that no observational
evidence of this ejection is seen suggests that the ejected material
becomes transparent on a timescale less than the 0.25~s bin time of
our time-resolved spectra.

Ejection of the accreted hydrogen likely requires that the accreted
material first becomes highly stratified.
In a well-mixed atmosphere, the H and He nuclei are efficiently coupled to
the electrons (on which the radiation forces act) through Coulomb
colisions, thus preventing any separation of the elements during the
radius expansion episode.  Variation in the H-fraction with depth 
may arise via a number of mechanisms, such as steady
H burning between the bursts or via convective mixing of deeper
material during the previous X-ray bursts. However, as we show below,
the properties of the X-ray bursts from \src\ place strong constraints on
these mechanisms.

In order to make 
a rough estimate for the requirements of
the ejection of the H layer during a burst, we assume that
all the radiation (emitted at $\simeq L_{\rm Edd,H}$) during the
radius-expansion episode is imparted as kinetic energy to the H layer.
Since the H layer needs to be ejected
in a short time $t_e\lesssim 1$~s, the total energy available is simply
$t_e$\leddh$\simeq 2\times 10^{38}$ergs. This amount of energy 
can unbind a layer of pure hydrogen of total mass 
\begin{eqnarray}
m_{\rm H} & \simeq & \frac{L_{\rm Edd,H}t_eR_{\rm NS}}{GM_{\rm NS}} \nonumber \\
& = &
     10^{18}\left(\frac{L_{\rm Edd,h}t_e}{2\times 10^{38}~{\rm ergs}}\right)
     \left(\frac{R_{\rm NS}}{10^6~{\rm cm}}\right) \nonumber \\
     & & \times\ \left(\frac{M_{\rm NS}}{1.4~M_\odot}\right)^{-1}~{\rm g},
\end{eqnarray}
which extends down to a column depth of
\begin{eqnarray}
y_{\rm H} & \simeq &
      10^{6}\left(\frac{L_{\rm Edd,h}t_e}{2\times 10^{38}~{\rm ergs}}\right)
     \left(\frac{R_{\rm NS}}{10^6~{\rm cm}}\right)^{-1} \nonumber \\
     & & \times\ \left(\frac{M_{\rm NS}}{1.4~M_\odot}\right)^{-1}~{\rm g~cm}^{-2}\;.
     \label{eq:y}
\end{eqnarray}
For the burst to be subsequently limited by the pure-He Eddington limit,
the material below this column depth must be practically hydrogen free. This 
degree of stratification is rather difficult to achieve on an accreting
neutron star, as we discuss below.

\cite{cb00} estimate the column depth at which hydrogen
runs out to be 
\begin{equation}
y_{\rm d}\simeq 7\times 10^{8} \left(\frac{\dot{m}}{0.01\dot{m}_{\rm Edd}}\right)
    \left(\frac{0.01}{Z_{\rm CNO}}\right)
    \left(\frac{X_0}{0.71}\right)~{\rm g~cm}^{-2}\;,
\end{equation}
where $\dot{m}$ is the mass accretion rate, $Z_{\rm CNO}$ is the mass
fraction of CNO nuclei, and $X_0$ is the mass fraction of hydrogen in
the accreting material.  For this column depth to be comparable to the
requirement derived in equation~(\ref{eq:y}), the accretion rate or the mass
fraction of CNO nuclei in \src\ have to be very different than what is
normally assumed. Depletion of hydrogen below a column depth of 
$\simeq 10^6$~gr~cm$^{-2}$ is possible, according to 
\cite{fl87},
if compressional heating is taken into account, the local accretion rate is
comparable to the Eddington limit, and the core temperature of the neutron
star is $\sim 10^{7}$~K. This is again a rather implausible combination, since
a high rate of accretion in a persistent source, such as \src, is inconsistent
with a cool neutron-star core
\cite[see, e.g.,][]{bbr98}.

Convective mixing of deeper material during the previous X-ray burst
also appears to be incapable of reducing the abundance of hydrogen at
columns larger than $\simeq 10^6$~gr~cm$^{-2}$. Indeed, the convective
zone in the simulations by 
\cite{woos03} reached up to column
depths that were at least two orders of magnitude larger than
required.  The properties of the bursts in \src\ thus strongly suggest
that there is an additional source of stratification which acts to
separate the accreted elements, and allow the majority of the accreted
H to be ejected during the brightest bursts.

In Paper A, we also explored in detail the
possibility that variations in \fpk\ could arise through systematic
instrumental or analysis-related biases. For example, we considered the
effect of deviations of the true emission spectrum from perfect
blackbodies.  Implicit in our estimation of the bolometric fluxes in
equation (\ref{flux}) is a correction to the flux measured in the PCA
bandpass; this correction adds around 7\% to the peak 2.5--20~keV PCA flux
of radius expansion bursts.  Should the emitted spectrum deviate
significantly from a blackbody, equation (\ref{flux}) will not give the
correct bolometric flux, potentially giving rise to spurious bolometric
flux variations.
As with \osrc, variations in the flux contribution from outside the PCA
band of many times the typical bolometric correction would be required to
account for the \highvar\% variation in the peak flux of the majority of
the radius expansion bursts from \src. We consider this unlikely.
It is even less likely that such effects could account for the
factor of $\approx 2$ variation we observe in the peak fluxes of all the
PRE bursts from \src.
Similarly, while variations in the persistent (non-burst) emission could
in principle give rise to spurious scatter in the peak burst fluxes, the
variation would have to be several times the total persistent flux level
in the 2.5--25~keV band to account for the peak flux variation we observe.

Although we found no evidence for quasi-periodic
variation in the peak fluxes of bursts from \src, or correlations between
the peak PRE burst flux and fluence, we cannot completely rule out
reprocessing as a 
contributing factor to the \highsig\ peak flux variation of the bright PRE
bursts.  The lack of
detectable quasiperiodic variations in the peak burst fluxes could result
from inadequate sampling provided by the burst times, or possibly the
intrinsic variability is aperiodic. As for the lack of a correlation
between \fluen\ and \fpk, the burst profiles from \src\/ were much more
variable than those of \osrc, so that other (possibly related) variations
between these properties could perhaps mask an underlying correlation
resulting from reprocessing.

An alternative explanation for the low peak flux PRE bursts is that
they arise from ignition of material confined to some fraction of the
neutron star.
Some authors have suggested that such confinement may occur as a result of
the influence of the neutron star's magnetic field on the fuel layer, or
as a consequence of rapid rotation
\cite[e.g.][]{szt85}. While recent modelling indicates that rotation may
instead enhance spreading \cite[]{slu02}, variations in the properties of
bursts with $\dot{M}$ from some sources suggest uneven distribution of
fuel over the neutron star \cite[e.g.][]{bil00}.  
This possibility
may also help to explain the observation of double- \cite[]{szt85} and even
triple-peaked \cite[]{vp86} bursts from \src.
We also observed four double-peaked bursts in the \xte\ sample.
Confinement of the burst fuel could result in weaker radius-expansion
bursts if the local Eddington limit was reached, but only over a fraction
of the neutron star surface.  In that case, expansion of the photosphere
would still occur, at much lower total luminosity.  If the expansion of
the burning region of the photosphere went to sufficiently large radii
that it appeared isotropic to a distant observer, the result might be a
burst which appeared similar to a normal radius
expansion burst, but at a significantly lower peak flux.
This explanation appears inconsistent with the observed radii in the tail
of the faint PRE bursts, which are in excess of that of some of the
brighter PRE bursts. It would also require that the factor of 1.7 between
the peak fluxes of the two groups of PRE bursts was a coincidence.
We note that several other explanations have been proposed to
explain the bursts with multiple peaks \cite[]{fuji88b,mz92}.

Previous estimates of the distance to \src\ at between 6--7~kpc relied on the
identification of the peak flux of the PRE bursts as \leddhe\
\cite[]{eb87}, in addition to a gravitational redshift measured from
absorption features in the burst spectra \cite[]{waki84}.
While absorption features have now been detected in burst spectra from
other sources using more modern, high-resolution spectroscopic instruments
\cite[]{cott02}, it is widely thought that the earlier detections using
proportional counters were more likely to arise from instrumental effects.
Without the absorption line measurement, the analysis of \cite{eb87}
likely cannot substantially constrain the distance to \src.
Despite the increase in measurement precision for peak burst fluxes with
\xte, the dominant uncertainty for distance estimates to most bursting
sources remains the atmospheric composition, i.e. the value of $X$ in
equation \ref{ledd}. With the detection of PRE bursts that
apparently reach either \leddh\ or \leddhe, the compositional question in
\src\ is resolved, and we can make distance estimates using either group
of bursts.  On correcting the observed fluxes for the gravitational
redshift at the surface of a $1.4M_\odot$ NS with $R=10$~km, the fainter
PRE bursts (which we assume reach \leddh) lead to a distance estimate of
$5.95\pm0.12$~kpc.  The distance estimate assuming that the brighter PRE
bursts reach \leddhe\ instead is fully consistent with this value at
\meandist, where the uncertainty arises from the standard deviation of the
peak fluxes for this larger group. From the faintest of this group of
bursts we can derive a maximum distance of \maxdist, or more
conservatively (for a $2M_\odot$ NS) \conupper.

\acknowledgments

This research has made use of data obtained through the High Energy
Astrophysics Science Archive Research Center Online Service, provided by
the NASA/Goddard Space Flight Center.  This work was supported in part by
the NASA Long Term Space Astrophysics program under grant NAG 5-9184.

\clearpage

\clearpage

\end{document}